\documentclass[fleqn,usenatbib]{mnras}

\usepackage{newtxtext,newtxmath}
\usepackage[T1]{fontenc}

\DeclareRobustCommand{\VAN}[3]{#2}
\let\VANthebibliography\thebibliography
\def\thebibliography{\DeclareRobustCommand{\VAN}[3]{##3}\VANthebibliography}

\newcommand{\varindex}{{$varindex$}}
\newcommand{\msol}{M$_\odot \,$}

\newcommand{\teff}{T$_{\rm eff}$}
\newcommand{\logg}{$\log(g)$}

\usepackage{graphicx}	% Including figure files
\usepackage{amsmath}	% Advanced maths commands
\title[Variable hot subdwarfs from ZTF]{Discovery of periodic hot subdwarf variables through a systematic search in Zwicky Transient Facility data}

\author[K. Wang et al.]{
Kevin Wang,$^{1,2}$\thanks{E-mail: kevinwangisme@gmail.com}
Thomas Kupfer,$^{2}$
and Brad~N.~Barlow$^{3}$
\\
$^{1}$Department of Computer Science, Princeton University, Princeton, NJ 08544, USA\\
$^{2}$Department of Physics and Astronomy, Texas Tech University, PO Box 41051, Lubbock, TX 79409, USA\\
$^{3}$Department of Physics and Astronomy, High Point University, High Point, NC, USA
}

\date{Accepted XXX. Received YYY; in original form ZZZ}

\pubyear{2015}

\begin{document}
\label{firstpage}
\pagerange{\pageref{firstpage}--\pageref{lastpage}}
\maketitle

\begin{abstract}
We conduct a systematic search for periodic variables in the hot subdwarf catalogue using data from the Zwicky Transient Facility. We present the classification of 67 HW Vir binaries, 496 reflection effect, pulsation or rotation sinusoids, 11 eclipsing signals, and 4 ellipsoidally modulated binaries. Of these, 486 are new discoveries that have not been previously published including a new mass-transferring hot subdwarf binary candidate. These sources were determined by applying the Lomb-Scargle and Box Least Squares periodograms along with manual inspection. We calculated variability statistics on all periodic sources, and compared our results to traditional methods of determining astrophysical variability. We find that $\approx60$\% percent of variable targets, mostly sinusoidal variability, would have been missed using a traditional \varindex\,cut. Most HW Virs, eclipsing systems and all ellipsoidal variables were recovered with a \varindex\,$>0.02$. We also find a significant reddening effect, with some variable hot subdwarfs meshing with the main sequence stripe in the Hertzprung-Russell Diagram. Examining the positions of the variable stars in Galactic coordinates, we discover a higher proportion of variable stars within $|b|<25^\circ$ of the Galactic Plane, suggesting that the Galactic Plane may be fertile grounds for future discoveries if photometric surveys can effectively process the clustered field.

\end{abstract}


%%%%%%%%%%%%%%%%%%%%%%%%%%%%%%%%%%%%%%%%%%%%%%%%%%

%%%%%%%%%%%%%%%%% BODY OF PAPER %%%%%%%%%%%%%%%%%%

\section{Introduction}

Most hot subdwarf B stars (sdBs) are core helium burning stars with masses around 0.5\,M$_\odot$\,and thin hydrogen envelopes (\citealt{heb86,heb09,heb16}). A large number of them are in tight binaries with orbital periods $<10$\,days \citep{nap04a,max01}. For these short--period sdB binaries, common envelope (CE) ejection is the only likely formation channel. One possible scenario is that two main sequence stars (MS) evolve in a binary system. The more massive one will evolve faster to become a red giant. Unstable mass transfer from the red giant to the companion will lead to a CE phase. Due to frictional forces, the two stellar cores lose orbital energy and angular momentum, which leads to a shrinkage of the orbit. The lost energy is deposited into the envelope, which spins up and ejects from the system. If the red giant core reaches the mass required for the core-helium flash before the envelope is lost, a binary consisting of a core-helium burning sdB star and a main sequence companion is formed. In another possible scenario, the more massive star evolves to become a white dwarf (WD) through either a CE phase or stable mass transfer onto the less massive companion. After that the less massive star evolves to become a red giant, and unstable mass transfer will lead to a CE. Once the envelope is ejected, the red giant remnant starts burning helium, and a system consisting of an sdB with a WD companion is formed \citep{han02,han03}. 

Most sdB stars show different kinds of variability in their light curves due to binarity or pulsations. Compact sdB binaries with WD companions can show ellipsoidal deformation in their light curves (e.g. \citealt{kup22,kup20a,kup20,pel21,gei13,blo11}). Compact sdB binaries with low-mass main sequence star companions show quasi-sinusoidal variability due to the reflection effect, resulting from the extreme temperature difference and small separation distance between the cool low-mass companion and the hot sdB star. Eclipsing reflection effect systems are called HW Vir binaries (e.g. \citealt{sch15,sch19,sch22,bar21,bar22}. Very few sdBs are known to show variability due to rotation. An exception to that is SB\,290 which shows a rotational period of $\approx5.5$ hours indicating a rapidly rotating sdB \citep{bal19}. \citet{pel20} found that many of the composite sdB binaries show small amplitude variations in their light curves, originating from spots on the surface of the cool companions.

Amongst the hot subwarf pulsators, two types of multi-periodic pulsators have been discovered, both with generally milli-mag amplitudes up to occasionally a few percent \citep{ost10}. On the hotter side (\teff$\gtrsim28,000$\,K) are the V361\,Hya stars which are pressure mode (p-mode) pulsators with typical periods of a few minutes \citep{kil97}. On the cooler side (\teff$\lesssim28,000$\,K) are the V1093\,Her stars which are gravity mode (g-mode) pulsators with periods of 45 min to 2 hours \citep{gre03}. A handful of pulsating sdB stars, known as hybrid pulsators, exhibit both gravity and pressure modes and have temperatures at the boundary between these two classes. Only a few sdB pulsators with a dominant radial mode are known, which show photometric amplitudes of up to several percent \citep{ore04,bar10, kup19, kup21}. A new class of pulsating hot stars known as Blue Large-Amplitude Pulsators (BLAPs) was discovered by \cite{pie17}. BLAPs show similar effective temperatures (\teff) as the sdBs but lower surface gravities (\logg) and are an order of magnitude more luminous at $L\approx10^2-10^3$\,L$_\odot$ with pulsation periods between 20-40 minutes. Given their unusual location on the HR diagram, it has been proposed that BLAPs are low-mass ($M\approx0.3$\,\msol) helium-core pre-white dwarfs \citep{cor18, rom18}. Delta Scutii and SX Phe stars are main sequence stars that have similar pulsation periods, although they are typically redder than hot subdwarf stars.

The known population of hot subdwarf binaries arose from mostly serendipitous discoveries \citep[e.g.][]{kup15a,gei15a}. However, large scale optical time-domain surveys have opened a new window to study the variable sky, providing hundreds to thousands of epochs across the whole sky. Starting with the Sloan Digital Sky Survey (SDSS; \citealt{yor00}), a new generation of wide-field optical surveys has exploited new affordable CCD detectors to open the frontier of data-intensive astronomy. This allows for the study of stellar variability and binary stars on different time-scales across a wide range of magnitudes. In particular, surveys covering also low Galactic latitudes are well suited to study the Galactic distribution of photometric variable stars. Ground based surveys include the Optical Gravitational Lensing Experiment (OGLE, \citealt{sos15}), the Palomar Transient Factory \citep[PTF,][]{law09}, the Vista Variables in the Via Lactea (VVV, \citealt{sai12}), the All-Sky Automated Survey for Supernovae \citep[ASAS-SN,][]{sha14,jay18}, the steroid Terrestrial-impact Last Alert System \citep[ATLAS,][]{ton18,hei18} and most recently the Zwicky Transient Facility \citep[ZTF,][]{gra19,mas19}.

\begin{figure*}
    \centering
    \includegraphics[width=\textwidth]{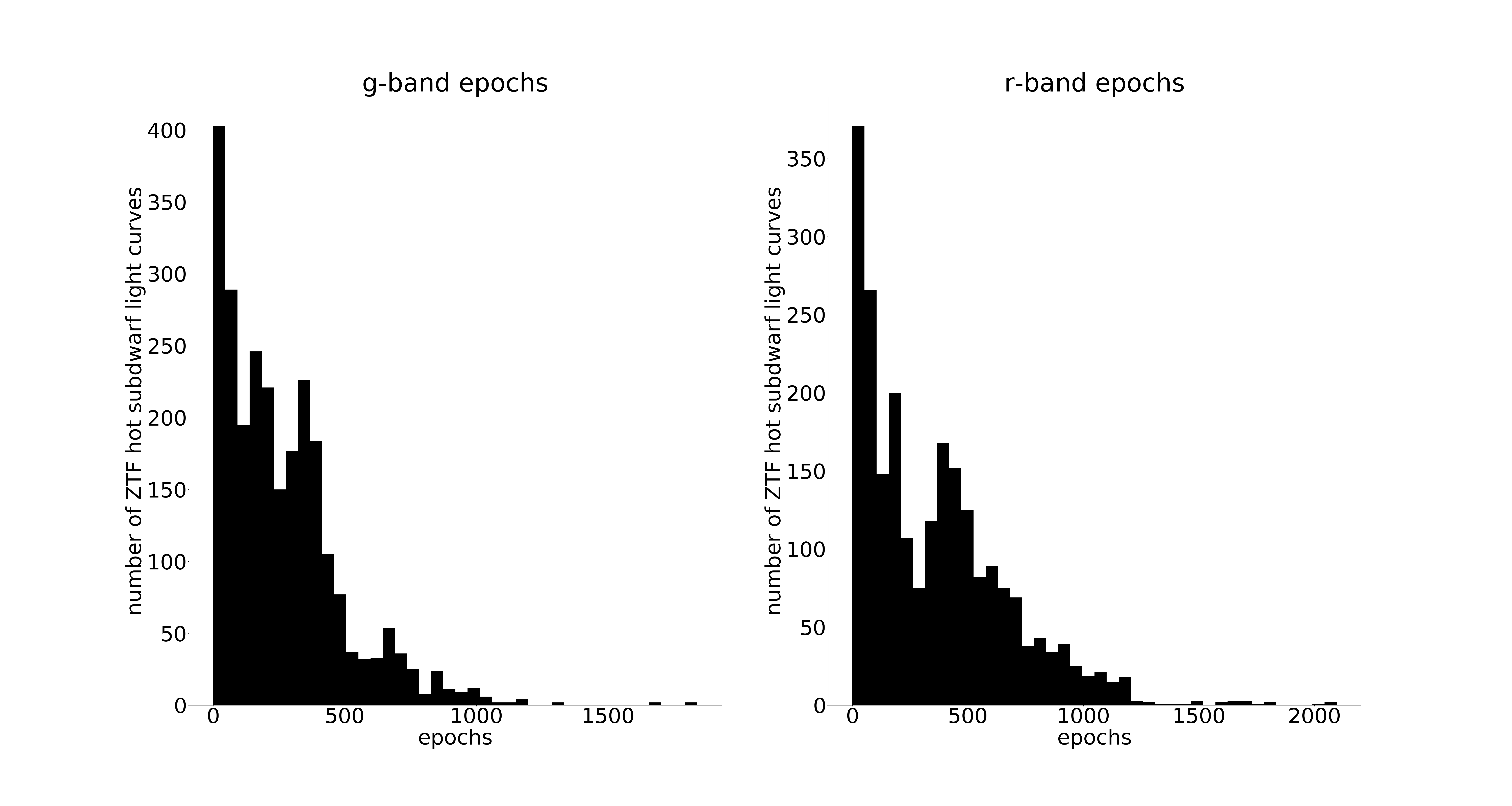}
    \caption{The distribution of the number of ZTF-$r$ and ZTF-$g$ epochs of our candidates from the hot subdwarf catalogue.}
    \label{fig:g-r_epochs}
\end{figure*}

Large populations of hot subdwarfs allow us to study population properties such as period and mass distributions \citep{sch22} which can be compared to theoretical predictions from binary evolution \citep{han02,han03, cla12,zha21,ge22}. Additionally, large numbers of known systems allow us to develop tools to distinguish between different types of hot subdwarf variables if the data quality and cadence allows it \citep{bar22}. In this paper we present a search for new variable hot subdwarf binaries, pulsators, and rotators in the catalogue from \citet{gei19} using data from ZTF. As part of ZTF, the Palomar 48-inch (P48) telescope images the sky every clear night conducting several surveys including a Northern Sky Survey with a 3-day cadence, as well as smaller surveys such as a 1-day Galactic Plane survey and simultaneous observations of the Northern TESS sectors \citep{gra19,bel19,bel19a,roe19}. Image processing of ZTF data is described in full detail in \citet{mas19}. ZTF has collected hundreds of photometric epochs over the last three years across the Northern hemisphere. 

\section{Methods and Results}

\subsection{Hot subdwarf catalogue and light curve extraction}

\citet{gei19} compiled a catalogue of 39,800 hot subdwarf candidates, which serve as the source of targets for our periodic variable search. The hot subdwarf candidates were chosen from stars in Gaia DR2 \citep{gai16,gai18} based on colour, absolute magnitude, and reduced proper motion cuts, which were derived by examining the population of hot subdwarfs in a previous catalogue \citep{gei17}. The catalogue aims to be correct rather than complete and the authors argue that the catalogue is fairly complete up to $\approx$1.5\,kpc, except in the direction of the Galactic plane and Magellanic clouds. In particular the number of sources in the Galactic plane is limited due to crowding and reddening. To account for composite sdB binaries, sources closer to the main sequence were included as well, and it is estimated that about 10\% of the catalogue might be contaminated by cooler stars which are not hot subdwarfs \citep{gei19}. 

The objects selected from the catalog of hot subdwarf candidates \citep{gei19} contain \textasciitilde5600 known hot subdwarfs listed by \citep{gei17}. A recent update has been given by \citet{2022A&A...662A..40C} listing 6616 spectoscopically confirmed objects, the vast majority being hot subdwarfs. We cross matched the ZTF sample of variable hot subdwarfs presented here with the sample of spectroscopically confirmed hot subdwarfs in \citet{2022A&A...662A..40C}. This is shown in the last column of Tab.\,\ref{tab:periodic_variables} where we indicate whether the newly discovered variable has been spectroscopically classified as hot subdwarf or not. In the latter case it remains a candidate.% Similarily, some of the hottest (unreddened) stars may be misclassified and are actually white dwarfs. A cross-match with the catalog of WDs (2019MNRAS.482.4570G) may show this, although we'd expect very few. 

%\subsection{Light curve extraction}
We extracted the ZTF light curves for the hot subdwarf sources from the catalogue using the Python module ztfquery\footnote{\url{https://github.com/MickaelRigault/ztfquery}} \citep{rig18} which queries the ZTF database and returns light curves for each object. We pulled light curves from ZTF DR6, which contains all public data obtained between March 2018 and April 2021 and private survey data obtained between March 2018 and December 2019. The private surveys include observational programs awarded by Caltech and performed by the ZTF collaboration, including a high cadence survey of the Galactic Plane \citep{kup21a}. Only objects with declinations greater than $-32^\circ$ and epochs greater than 30 were selected, resulting in 18,453 targets. Times were converted to barycentric Julian date (BJD) using the astropy module \citep{astpy13,astpy18}, and normalized fluxes ($f_{{\rm norm}}$) were derived from the magnitudes for each light curve ($m_1$) with the average magnitude ($m_{{\rm avg}}$) across the light curve as the reference point. To increase the number of epochs for our period finding algorithm, we combined ZTF-$r$ and ZTF-$g$ normalized flux data in temporal space before doing the periodicity analysis. We note that many variable hot subdwarfs exhibit wavelength--dependent amplitudes, and thus their light curves could have different shapes in the ZTF-$r$ and ZTF-$g$ filters. However, our project focused on identifying periodic variables and their periods --- not on measuring their amplitudes or light curve morphologies accurately --- and so combining together the ZTF-$r$ and ZTF-$g$ data only enhances our ability to uncover variables. Our list of hot subdwarf candidates have an average of 345 epochs in the ZTF-$r$ and 282 epochs in the ZTF-$g$. The ZTF epoch distributions in ZTF-$g$ and ZTF-$r$ for our candidates are shown in Fig.\,\ref{fig:g-r_epochs}.

%\begin{equation}
%f_{{\rm norm}} = 10^{-0.4(m_1-m_{{\rm avg}})}
%\end{equation}

%To estimate the uncertainty in the flux ($\sigma_f$) from the magnitude uncertainty ($\sigma_m$) we start with the equation to convery magnitude to flux equation, with $m$ being the magnitude and $f$ the flux:

%\begin{equation}
%m = -2.5\log_{10}f + c
%\end{equation}

%Taking the derivative, we get:

%\begin{equation}
%\frac{dm}{df} = \frac{-2.5}{\ln10f} \approx \frac{-1.09}{f}
%\end{equation}

%Since:

%\begin{equation}
%\sigma_m = \frac{dm}{df}\sigma_f
%\end{equation}

%We estimate:
%\begin{equation}
%|\sigma_m| \approx |\frac{-1.09}{f}\sigma_f| \approx \frac{\sigma_f}{f}
%\end{equation}

\begin{figure*}
    \centering
    \includegraphics[width=\textwidth]{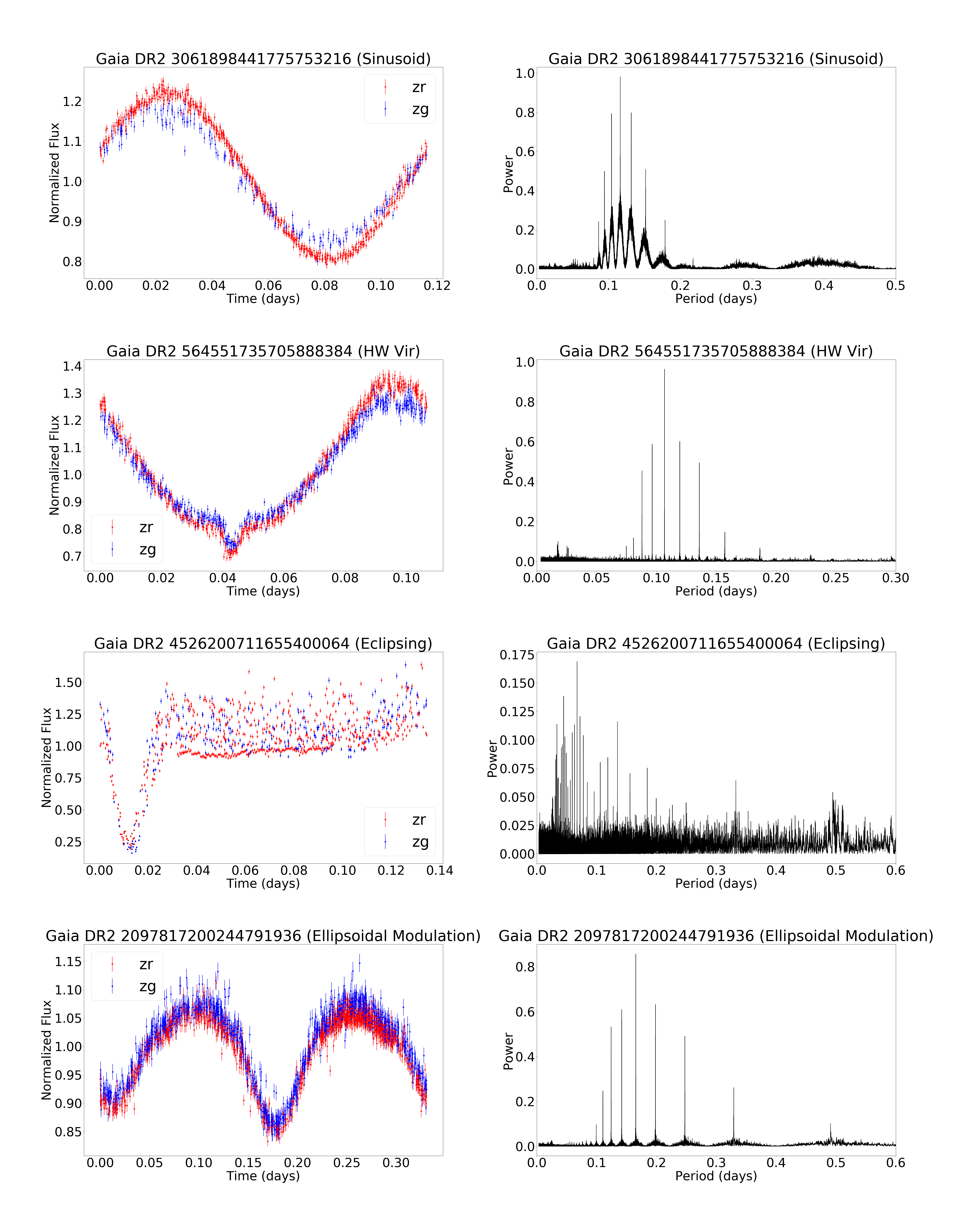}
    \caption{Examples of a sinusoidal system, an HW Vir, an eclipsing system, and an ellipsoidal binary. On the left, phase folded light curves at the best period, and on the right, the corresponding Lomb-Scargle Periodogram. Although lomb-scargle was always applied with a period range of 0 - 3.0 days, the periodograms on the right are shown over a narrower range of periods so as to zoom into the areas containing peaks.} 
    \label{fig:examples}
\end{figure*}

\begin{figure*}
    \centering
    \includegraphics[width=\textwidth]{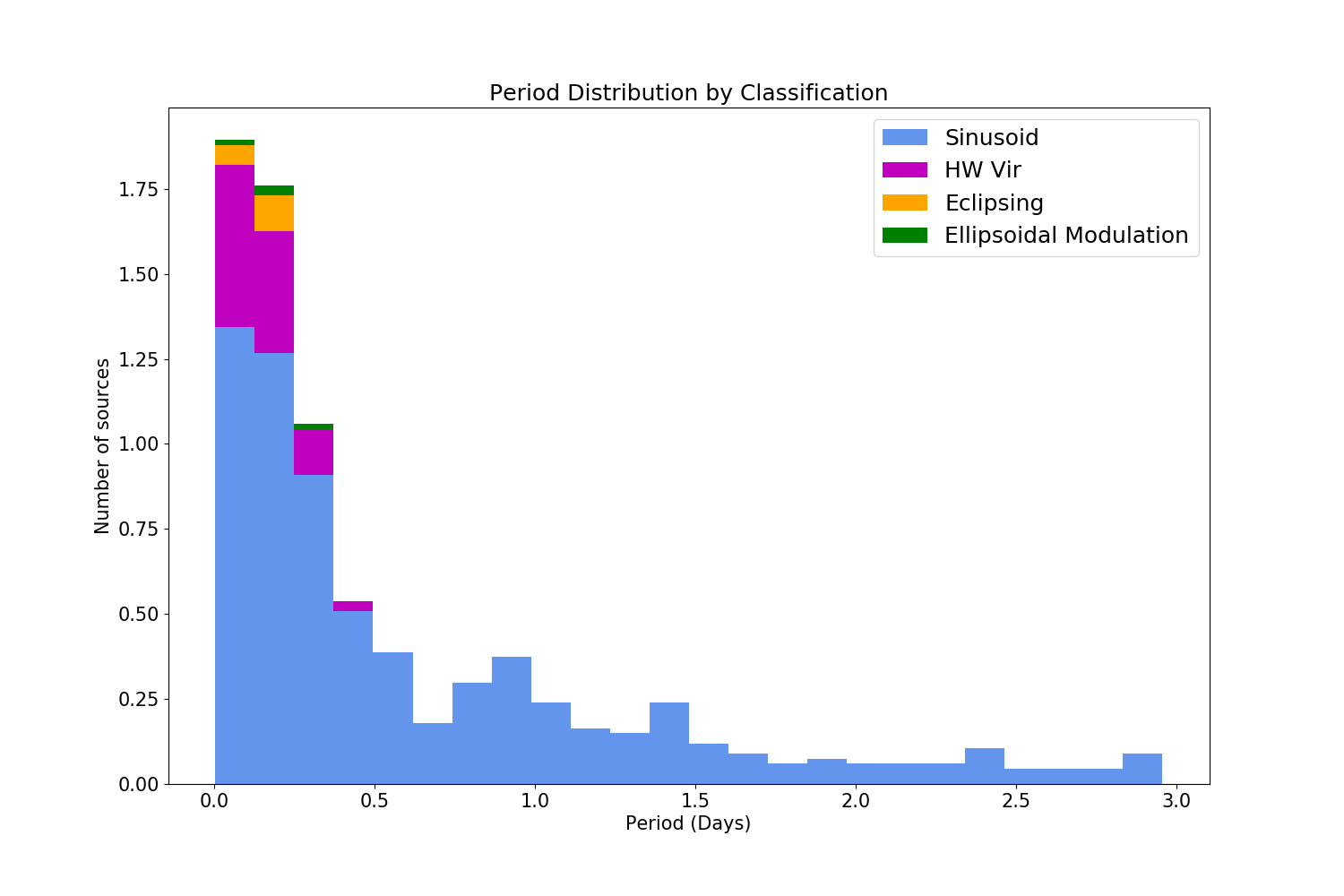}
    \caption{Period distributions for variables classified as sinusoidal, HW Vir, eclipsing, and ellipsoidal modulation.}
    \label{fig:period_dist}
\end{figure*}

\subsection{Periodicity analysis }
To find physical periods in the data, we used the Lomb-Scargle periodogram \citep{Lom76,sca82} as the primary period-finding method, since most of the anticipated stellar variability is quasi sinusoidal. We expect to find mostly reflection effect systems, HW\,Vir binaries, eclipsing sources and pulsators with periods $\leq$ a few days. On the lower end of the period search range are pulsators whose periods can go as low as a few minutes. On the high end of the period search range are reflection effect binaries and binary eclipses, which can go up to a few days. Because we mainly focus on short period binaries and pulsators we searched for variability from 5 minutes to 3 days with the Lomb-Scargle periodogram, using 10 grid points across each significant periodogram peak in our search. We employ the Lomb-Scargle periodogram from the astropy module \citep{astpy13, astpy18}. For each source, we calculated both the max power and the the Signal Detection Efficiency (SDE) as shown in equation\,\ref{equ:sde}.

\begin{table*}
\begin{tabular}{|p{3.4cm}||p{1.7cm}|p{1.7cm}|p{1.7cm}|p{1.7cm}|p{1.7cm}|p{1.7cm}|p{1.6cm}}
 \hline
 \multicolumn{8}{|c|}{Periodic Variables} \\
 \hline
 Gaia DR2 ID & RAJ2000  & DEJ2000 & period & Gaia G-band & classification & previously & spectral\\
   & [deg] & [deg] & [days] & [mag] & & known & classification\\
 \hline
3061898441775753216&112.58808&-2.108157&0.1163&15.54&sinusoid&0&N/A\\
564551735705888384&4.154748&78.921942&0.1068&16.54&HW Vir&0&N/A\\
4112644491182137984&256.733194&-24.984093&0.0949&14.24&sinusoid&0&N/A\\
2089529322115136128&299.36842&53.534251&1.3631&14.95&sinusoid&0&sd\\
6850608253147209088&302.706481&-24.978347&1.2521&16.48&sinusoid&0&N/A\\
4194507602931516672&293.983632&-8.86162&0.1523&15.69&sinusoid&0&N/A\\
2182023160826160000&311.659035&51.793221&0.0896&15.23&HW Vir&1&sdB\\
3010515995663287424&84.355931&-10.765187&1.0453&16.42&sinusoid&0&N/A\\
1934910499455947904&346.626862&44.313543&0.0879&14.29&sinusoid&1&sdB\\
4299127989744823424&300.870726&8.646449&0.0986&14.48&sinusoid&0&N/A\\
\hline
\end{tabular}
\caption{First ten rows of the table of periodic variables recovered from the hot subdwarf catalogue. Previously known systems are marked with 1 and previously unknown systems are marked with 0.  The last column was done with a cross match with \citep{2022A&A...662A..40C}, which lists 6616 sdBs and their spectral classification. 115 out of 578 periodic variables are in \citet{2022A&A...662A..40C}'s catalog, and the other 463 periodic variables have their spectral classification labeled as "N/A". The full table is available online. }
\label{tab:periodic_variables}
\end{table*}

\begin{equation}\label{equ:sde}
SDE_{LS} = \frac{P(f_{max}) - \langle P \rangle}{sd\langle P \rangle}
\end{equation}
where P is the Lomb-Scargle periodogram function, $\langle P \rangle$ is the mean of the periodogram powers, and $sd\langle P \rangle$ is the standard deviation of the periodogram powers. 

To search for physical periodic sources which do not show quasi sinusoidal variability but just eclipses, we also applied the Box Least Squares method \citep{kov16} as an auxiliary periodogram as a reference while classifying the light curves. Box Least Squares is also imported from astropy. While the vast majority of categorizations were made referring to the Lomb-Scargle periodogram, a few eclipsing systems were detected by BLS only; eclipsing signals are transit-like, whose shape is similar to a box shape, rather than a sinusoid, the fitting mechanism that Lomb-Scargle employs. We used a period range from 20 minutes to 3 days for BLS since eclipsing hot subdwarf candidates usually have periods ranging from 20-30 minutes up to a few days. We used 1 million evenly spaced periods with a duration estimate of 5 minutes. For each object, we made two phase folded plots: one at the peak period determined by the Lomb-Scargle and one at the peak period determined by BLS. There is no standard quantifiable evaluation mechanism to determine whether a peak in a Lomb-Scargle periodogram indicates true quasi sinusoidal variability. We chose not to utilize a false alarm probability cutoff because it assumes Gaussian distribution on the periodogram noise, which is unrealistic with ZTF's low number of epochs. Thus, we classify our targets manually. Because of the conservative cuts in the catalogue compilation from \citet{gei19} which already preclude completeness, we categorize our targets conservatively to ensure reliability of our classifications.

We first sorted the results by the power at the most likely period that the Lomb-Scargle periodogram determines. Because ZTF is a ground-based survey, it observes only at night time, causing data gaps which can significantly bias periodicity analysis to find periods of 1.0 or 0.5 days. As such, light curves which are determined to have a period of 1.0 or 0.5 days may not have those periods in actuality. Thus, we filtered out all objects which Lomb-Scargle determined to have a period around 1.0 or 0.5 days by applying empirically-derived crop ranges of (0.495 days - 0.52 days) and (0.99 days - 1.04 days). Then, for each source, we look for both eclipsing and quasi sinusoidal signals. Quasi sinusoidal signals are expected to consist mostly of pulsators and reflection effect systems. Rotation has also been detected in space-based data of rapidly rotating sdBs as well as from the cool companions in compsite sdBs \citep{pel20,bal19}. However, in both cases the variability amplitudes are below the detection limit of ground based surveys and as such we do not expect to detect rotation in our data. To identify eclipses, we manually compiled targets with transit-shaped dips. To identify physically real quasi sinusoidal signals, we manually derive empirical SDE cuts that ensure conservative classification---a light curve was only labelled as a quasi sinusoidal signal if it displayed a combination of high maximum power and high SDE value. Specifically, the derived cutoffs are as follows: for the 300 light curves with the highest max powers, sources with SDE greater than 15 were classified as sinusoids; for the next 700, sources with SDE greater than 18 were classified as sinusoids; for the next 500, sources with SDE greater than 28 were classified as sinusoids; for the next 1500, sources with SDE greater than 36 were classified as sinusoids; for the next 4000, sources with SDE greater than 50 were classified as sinusoids; for the rest, no sources were classified as sinusoids. Subsequently, we manually removed blending, flares, and other aperiodic variation that Lomb-Scargle picked up which were clearly not sinusoidal variation. To confirm our results, we verified by manual inspection that each of resulting phase folded light curves showed a sinusoidal structure. In the final step, we combined our sinusoidal and eclipsing light curves and classified each target to be in one of these classes: HW Vir binary, eclipsing binary, ellipsoidal modulation, sinusoidal, or none. We gave the label HW Vir to any light curve which had a sinusoidal reflection effect baseline with a transit at the trough of the sinusoid, and a weaker transit at the peak of the sinusoid. We classified eclipsing for any light curve with at least one transit and no reflection effect baseline. Ellipsoidal modulation was given to any light curve which showed a  sinusoidal shape with a sinusoidal dip followed by a shallower or deeper sinusoidal dip coming from gravity darkening. Sinusoidal was given to any light curve which showed a quasi sinusoidal shape with no eclipse and gravity darkening effect leading to different minima depth. All other light curves were given the classification of none. Fig.\,\ref{fig:examples} shows the phase folded ZTF light curve as well as the Lomb-Scargle periodogram as an example for each class. 

\subsection{Periodic variables and Gaia crossmatch}
Applying the methods described above, we find a total of 578 periodic variables with signals dominated by sinusoidal classification. We found 496 sinusoidal systems, 67 HW Virs, 11 eclipsing systems, and 4 ellipsoidal modulated systems. For each of the 578 periodic variables, we conducted a publication check through SIMBAD's reference query mode in early October of 2021. Typical previously published catalogs include \citet{gei11b,gei14,kup15a,hol17,sch19, sch22,sah20,rat20,bar21,krz22,dai22,sol22,bar23}. Of the 578 periodic variables, 435 sinusoids, 44 HW Virs, 5 eclipses and 2 ellipsoidal modulations are new classifications while the rest had been previously discovered. We provide a table of information about the 578 periodic variables, including their Gaia DR2 identifiers, their coordinates, their orbital period, their classification, whether the system has been previously published, and their spectral classification. The spectral classification was labeled by cross-referencing with \citet{2022A&A...662A..40C}, which catalogues the 6616 sdBs along with their spectral classification. From the 578 periodic variables, 115 are found in \citet{2022A&A...662A..40C}'s catalog, while the remaining 463 periodic variables are classified as "N/A". The first few rows are shown in Tab.\,\ref{tab:periodic_variables}. We will provide the full catalogue online. The Lomb-Scargle periodogram and the phase folded light curve is shown for one example of each type of periodic variability in Fig.\,\ref{fig:examples}. For each classification of the periodic variables, we plot the period distributions based on the maximum period that the Lomb-Scargle found---except for periodic variables which were either only detected by BLS or whose phase fold at the Lomb-Scargle period showed the period a harmonic of the true period; in both of these cases, we corrected the period accordingly to the true period. Fig. \ref{fig:period_dist} shows the period distribution for all four different types of variable stars. Most objects show periods well below one day and as short as a few minutes.

%\begin{figure*}
%    \centering
%    \includegraphics[width=\textwidth]{em_examples_newtitles.png}
%    \caption{Phase folded light curves of two ellipsoidal systems Gaia DR2 4200455960845144960 %and Gaia DR2 4082986745566106368. For both systems, one minimum is much deeper than the other, %indicating a possible eclipse from an accretion disc.}
%    \label{fig:em}
%\end{figure*}

\begin{figure*}
    \centering
    \includegraphics[width=\textwidth]{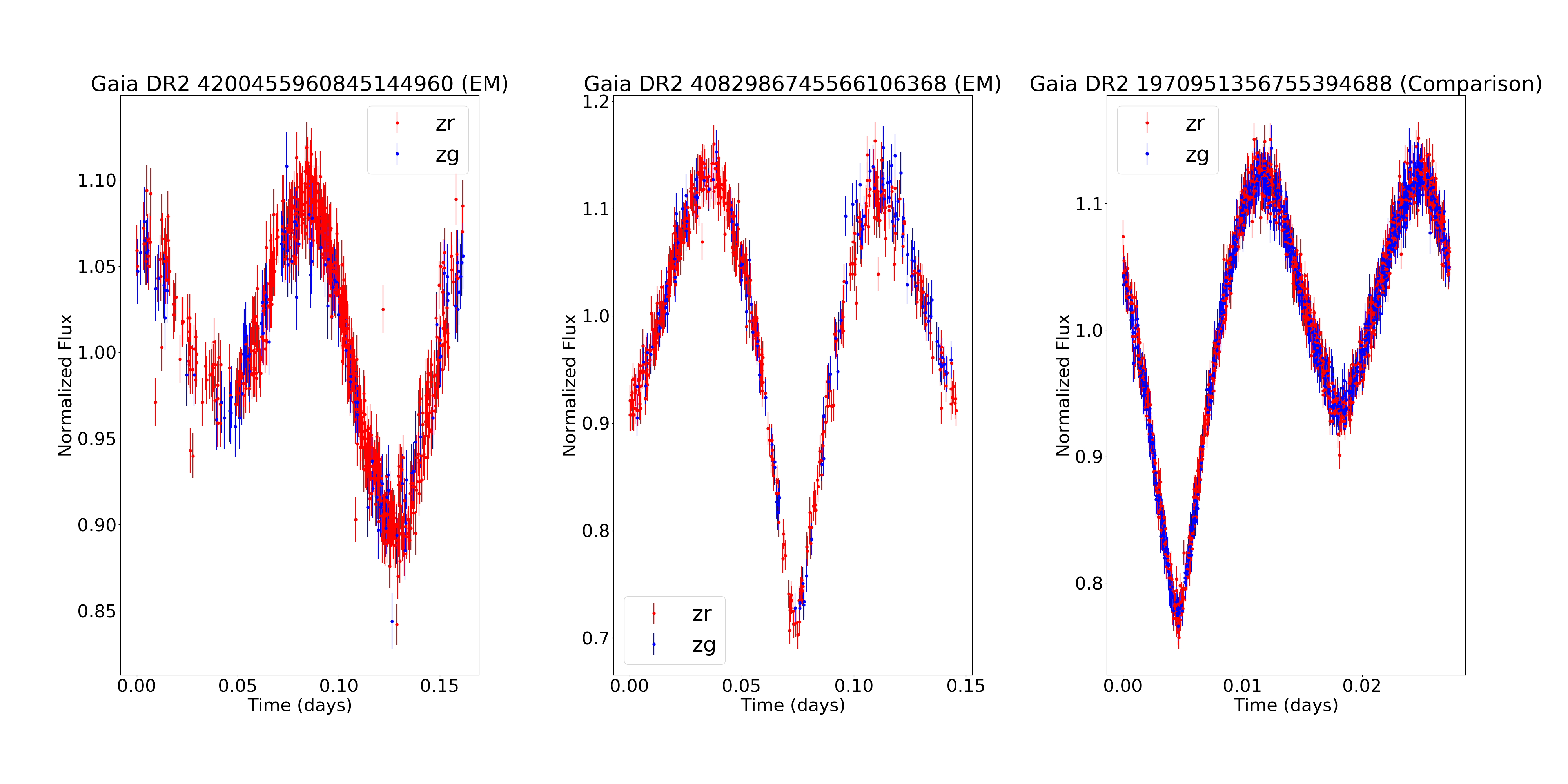}
    \caption{Phase folded light curves of two ellipsoidal systems Gaia DR2 4200455960845144960 and Gaia DR2 4082986745566106368. For both systems, one minimum is much deeper than the other, indicating a possible eclipse from an accretion disc. Gaia DR2 1970951356755394688 is known to be a mass accretion hot subdwarf star and is shown for comparison.}
    \label{fig:em}
\end{figure*}

\begin{figure*}
    \centering
    \includegraphics[width=\textwidth]{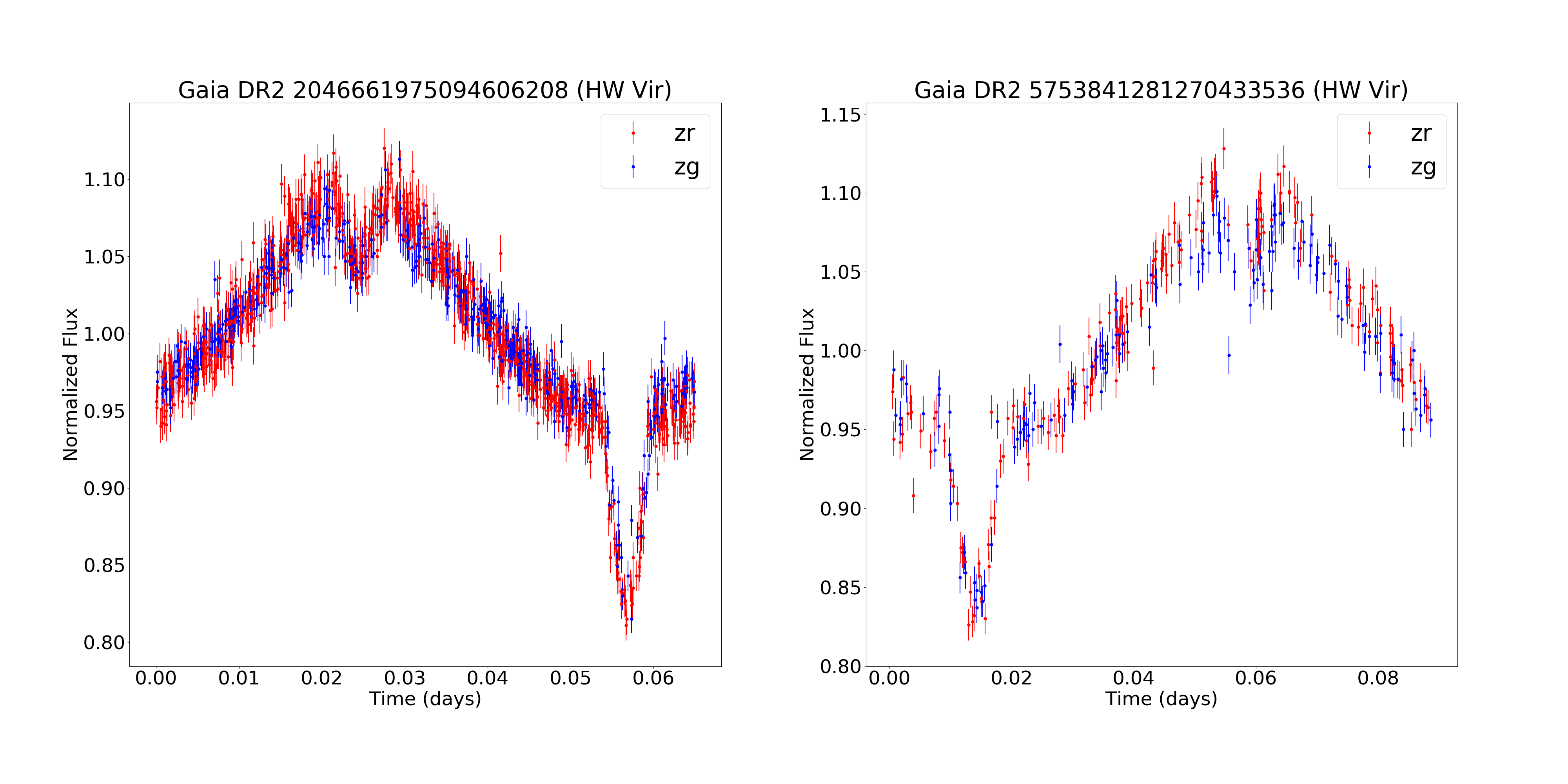}
    \caption{Phase folded light curves of two HW Vir systems Gaia DR2 2046661975094606208 and Gaia DR2 5753841281270433536. For both systems, a reflection effect is accompanied by a deep eclipse and its secondary eclipse, classifying these sources as HW Virs.}
    \label{fig:hw}
\end{figure*}

\begin{figure*}
    \centering
    \includegraphics[width=\textwidth]{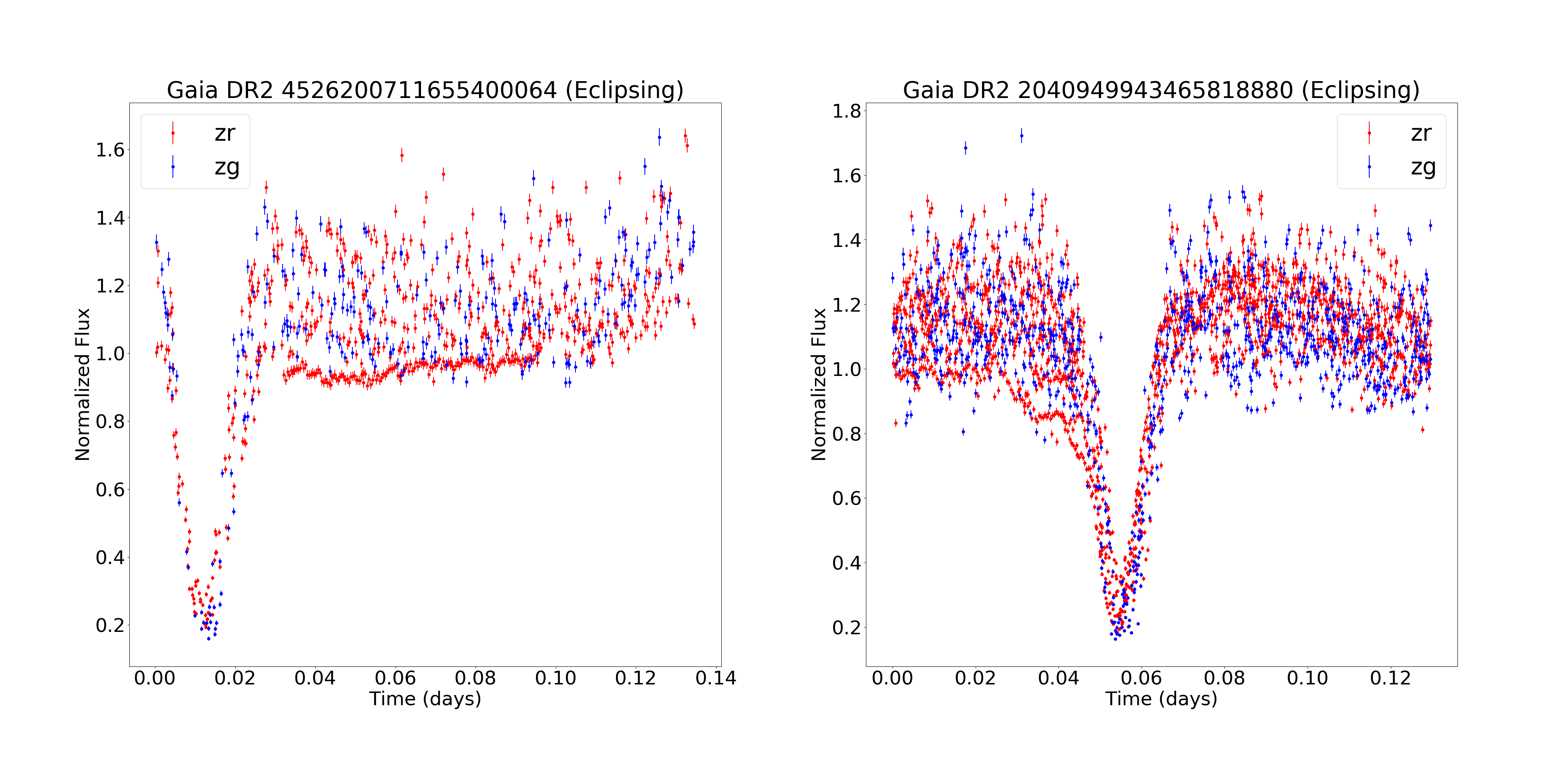}
    \caption{Phase folded light curves of two eclipsing systems Gaia DR2 4526200711655400064 and Gaia DR2 2040949943465818880. The high variability of the light curve indicate accretion, classifying these sources as cataclysmic variables.}
    \label{fig:ec}
\end{figure*}

\begin{figure*}
    \centering
    \includegraphics[width=\textwidth]{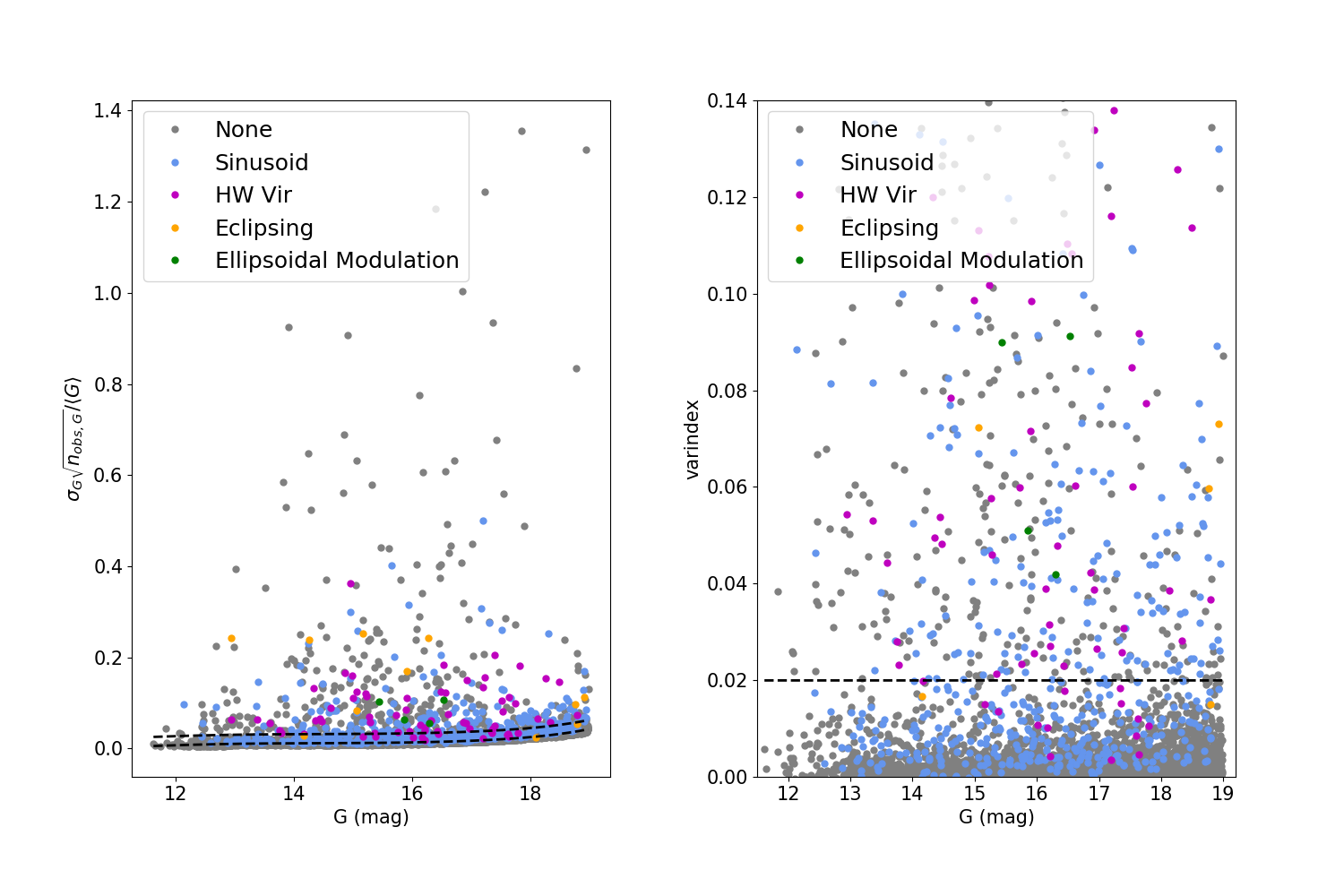}
    \caption{{\bf Left panel:} The 18,453 ZTF hot subdwarf candidates graphed on a normalized flux variation vs. magnitude plot. The lower black dashed line represents the fit of the main gray stripe determined by a low-polynomial regression after iterative outlier removal is done. The upper black dashed line represents the cut of 0.02 above the lower black dashed line fit. {\bf Right panel:}  The 18,453 ZTF hot subdwarf candidates plotted as  a function of \varindex. In previous studies systems with \varindex\,$>0.02$ are considered variable.}
    \label{fig:var}
\end{figure*}

%\subsection{Gaia crossmatch}

Using the coordinates of the targets in the catalogue, we cross-matched our selected targets with Gaia eDR3 catalogue \citep{gai16, gai21} through the interactive tool TOPCAT \citep{tay05}. Gaia is a space satellite which provides the parallax [$\varpi$] which can be converted to distances for more than 1 billion objects on the sky. The results from the crossmatch with Gaia data were combined with the periodogram results for each source. Because the Hertzsprung-Russell (H-R) diagram analysis relies on precise distance measurements, we differentiated between sources which had a parallax over parallax error ($\frac{\varpi}{\sigma_\varpi}$) larger than 10. To derive the distance we simply used $\frac{1}{\varpi}$ and no quality cuts were applied to ensure as complete a sample as possible.

\section{Discussion}

\subsection{Sinusoidal systems}
We would expect to find reflection effect systems, pulsators and sinusoidal variability due to rotation. These different types have distinct quasi sinusoidal variability. The reflection effect is symmetric about its peak and deviates from a sinusoidal shape, with the crests being slightly sharper and the troughs being flatter whereas radial mode pulsators typically show a sawtooth shape with a steep rise and a flat fall. In almost all cases the light curve shape does not allow to distinguish for certain between pulsators, rotators and reflection effect without follow-up observations. Another physical phenomenon affecting light curves is Doppler boosting, also known as beaming, which is a physical phenomenon that occurs when a source of light is moving at a high velocity; however, the effect is so small that space-based quality data is required for detection \citep{sch22}. Therefore we do not differentiate between our sinusoidal signals.

A possible way to distinguish between compact pulsators or compact rotators and reflection effect systems is the period. Reflection effect systems with an sdB component cannot reach orbital periods below 60\,min. We found 24 systems amongst our sinusoids with a period below 60\,min, indicating that they are compact pulsators or compact rotators. The shortest period of our sinusoidal systems is 0.00361 days or 5.20 minutes for Gaia DR2 1486741519792600448. Most of the systems have periods below $\approx$0.5\,days with very few systems with periods above 1 day. 

% Overall, though, we note that 24 of our sinusoids have an orbital period less than 60 min, strongly indicating that they are pulsators. The shortest period of our sinusoidal systems is 0.003614366 days or 5.20469 minutes.

\begin{figure*}
    \centering
    \includegraphics[width=\textwidth]{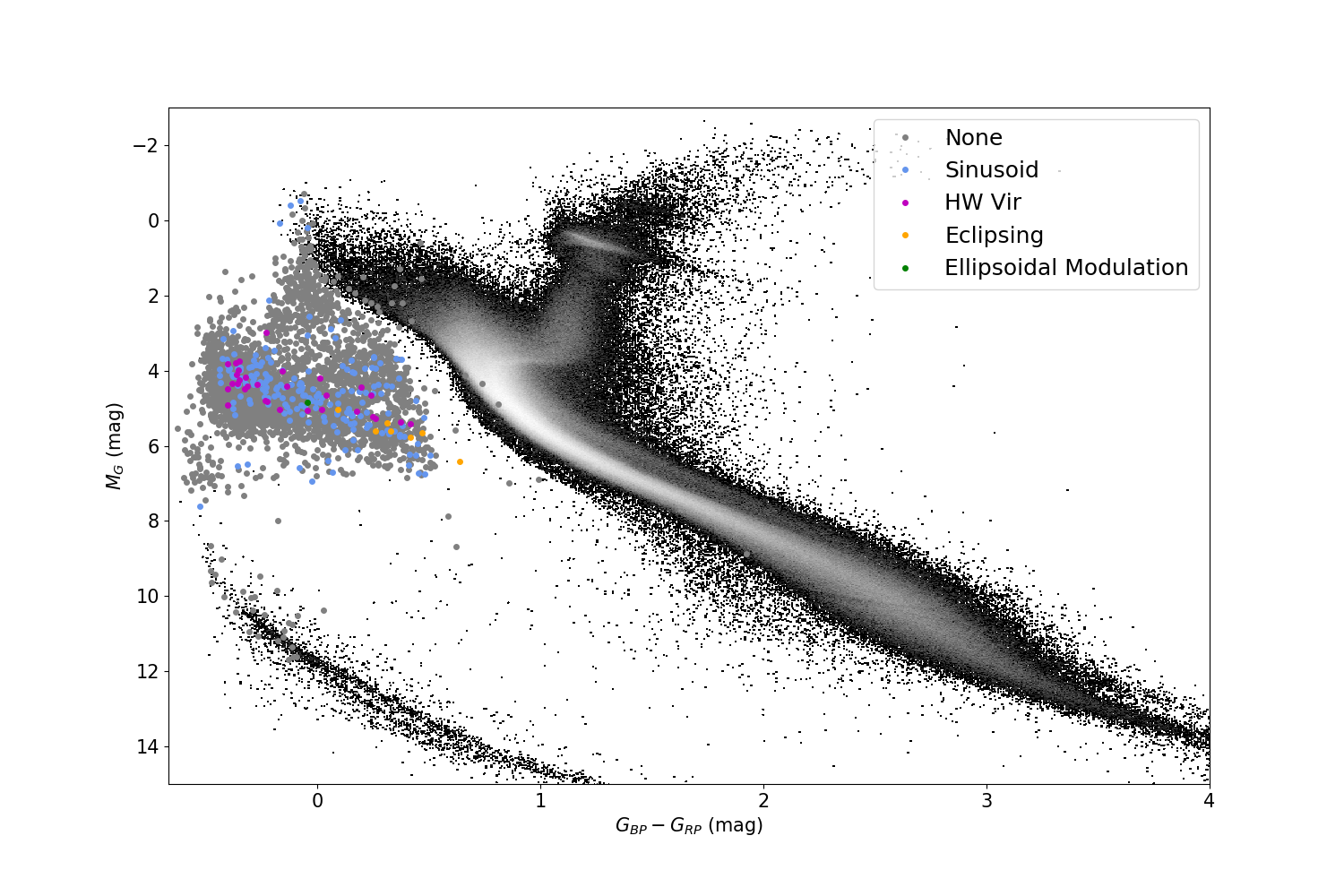}
    \caption{H-R diagram of hot subdwarf candidates with parallax precision better than 10\%. Catalogue sources are plotted with the gray and colored data points on the upper-left subdwarf region of the diagram. Other stars are shown in a heatmap based on density on a log scale. A reddening effect can be seen, as many of the variable subdwarfs are to the right of the traditional subdwarf region, meshing into the main sequence.}
    \label{fig:hr}
\end{figure*}

\begin{figure*}
    \centering
    \includegraphics[width=\textwidth]{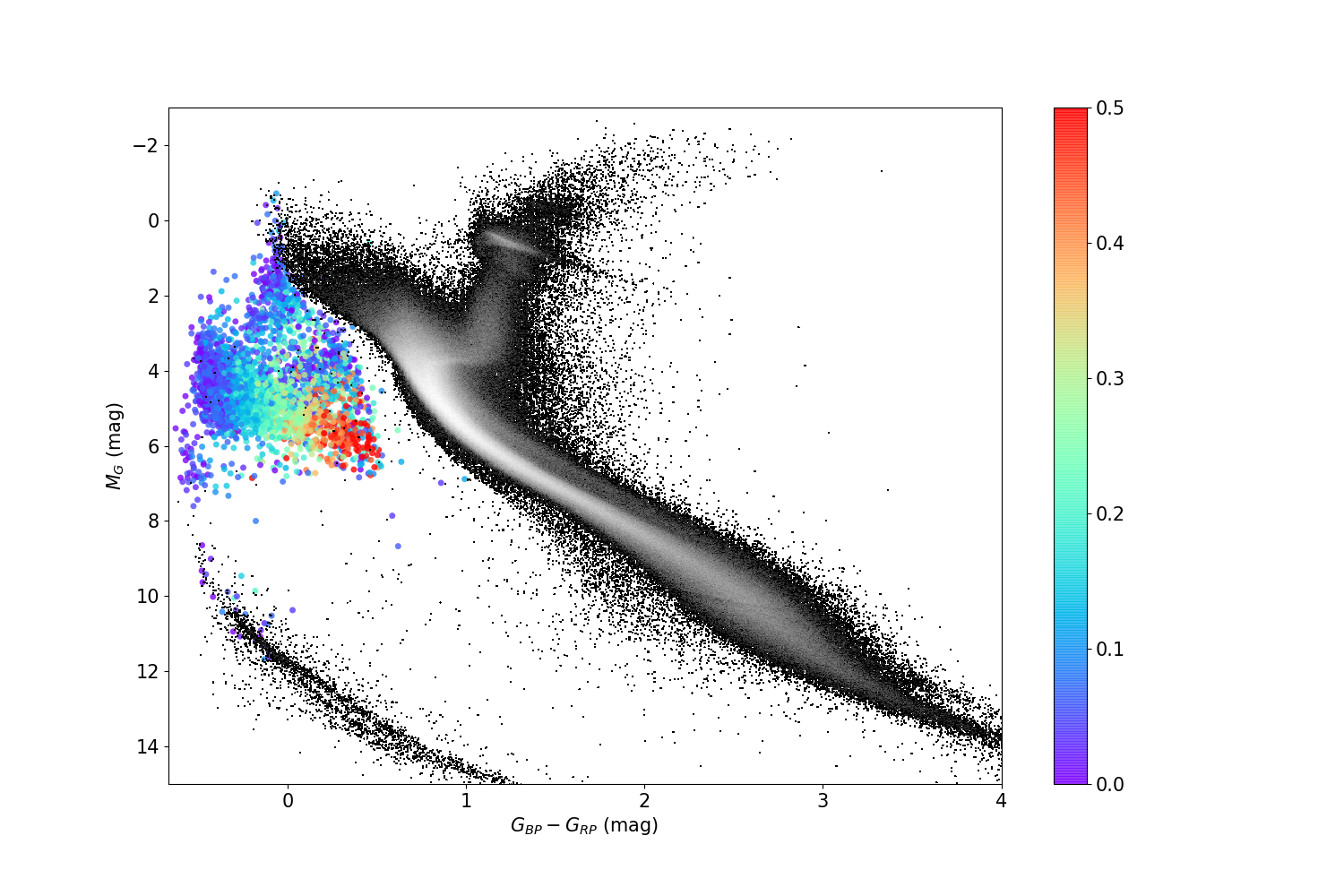}
    \caption{H-R diagram of hot subdwarf candidate reddening for sources with parallax precision better than 10\%, showing a reddening effect closer to the main-sequence. The colorbar represents the extinction code from \citet{gre19}.}
    \label{fig:reddening-heat}
\end{figure*}

\subsection{Ellipsoidal systems}
Ellipsoidally modulated systems are sdB binaries with white dwarf companions where the sdB is tidally deformed. The tidal deformation causes a sinusoidal variability but the light curves show two different minima during one orbit due to gravity darkening. Gravity darkening effects are less than a few percent and therefore can easily be missed in ground based data. Recently \citet{kup20, kup20a} discovered sdOB binaries with very large minima differences of $>10$\,\% and detailed modelling confirmed that both systems contain an accretion disc around the white dwarf which eclipses the sdOBs leading to larger differences in the minima typically observed with gravity darkening alone. Our search also contained ZTF\,J2130; a previously published mass transferring system \citep{kup20}.

Our search revealed two systems which show a similar light curve shape than the two known accreting systems. Gaia DR2 4200455960845144960 and Gaia DR2 4082986745566106368 show one minimum significantly deeper than the other. See Fig.\,\ref{fig:em} for the phase folded light curves. Because of the similarities between these systems and the previously published mass transferring systems, we postulate that Gaia DR2 4200455960845144960 and Gaia DR2 4082986745566106368 may also be accreting systems. This is in agreement with a study by \citet{li22} who present an analysis on Gaia DR2 4082986745566106368 showing that light curve models without an accretion disc do match the observed light curve shape. Notably, the orbital periods of these two systems, 3.89 hours and 3.50 hours, are much longer than the previously published systems which show orbital periods of 39 minutes and 56 minutes. We note that contact binaries show a similar light curve shape, however, contact binaries with 3-4 hour orbital period would consist of low-mass main sequence which is not consistent with the color of our objects.

\subsection{HW binaries}

HW Virs are eclipsing sdB binaries that show a reflection effect. They typically consist of an sdB/sdO with an M-dwarf or substellar companion \citep[e.g.][]{sch14,sch15,sch21}. In 2021, \citet{Enenstein_2021} reported the discovery of 26 new HW Vir systems from the \citet{gei19} catalog. Currently the largest sample was detected in the EREBOS survey which aims to statistically study the population of HW Vir binaries \citep{sch19,sch22,cor21}. They present a sample of 169 objects with more than half of them discovered in the OGLE survey.

Our search recovered 67 HW\,Virs, 23 of which were already known, including 13 HW\,Virs from the EREOBS sample. All HW\,Virs from our sample show the typical light curve with a deep primary eclipse a weak secondary eclipse and a strong reflection effect. Fig.\,\ref{fig:hw} shows two examples of these HW Vir systems that were not previously published. Most newly discovered systems have periods below 0.2 days and no system was detected with a period above 0.5 days which is consistent with results from the EREBOS sample. The most compact system in our sample is Gaia DR2 3339211934473498368 with a period of $89$\,minutes. The 44 new HW\,Virs discovered in this sample may be useful for population analysis of HW\,Vir binaries such as the one conducted by \citet{sch19} and \citet{sch22}.

\subsection{Eclipsing systems}
We identify eclipsing sources as sources which show deep eclipses without an obvious reflection effect. In our eclipsing population, over half of the systems have high variability outside the eclipse, where the light curve noise is significantly larger than the flux errors of the data points. suggesting that they may in fact be cataclysmic variables (CVs). Fig.\,\ref{fig:ec} shows two examples of these eclipsing systems that were not previously published. The significant flickering of the light curve is a strong sign of accretion. Additionally, we can infer that these systems may have a hot accreting white dwarf component, since the light from the donor stars are so faint, as indicated by the lack of a secondary eclipses. 

CVs show a large range of colors and absolute magnitudes and \citet{abr20} show that nearly all types of CVs fall within the same part of the Gaia color-magnitude diagram \citet{gei19} used for their hot subdwarf selection criteria. This includes nova--like CVs, dwarf novae, classical novae, and intermediate polars. Only WZ Sge and polars are (generally) outside these bounds and therefore it is not surprising that we find these systems in the \citet{gei19} catalogue. The eclipsing CVs will be useful for follow-up studies as the eclipses allow to constrain the inclination angle, a crucial ingredient for detailed paremeter studies.

\begin{figure*}
    \centering
    \includegraphics[width=\textwidth]{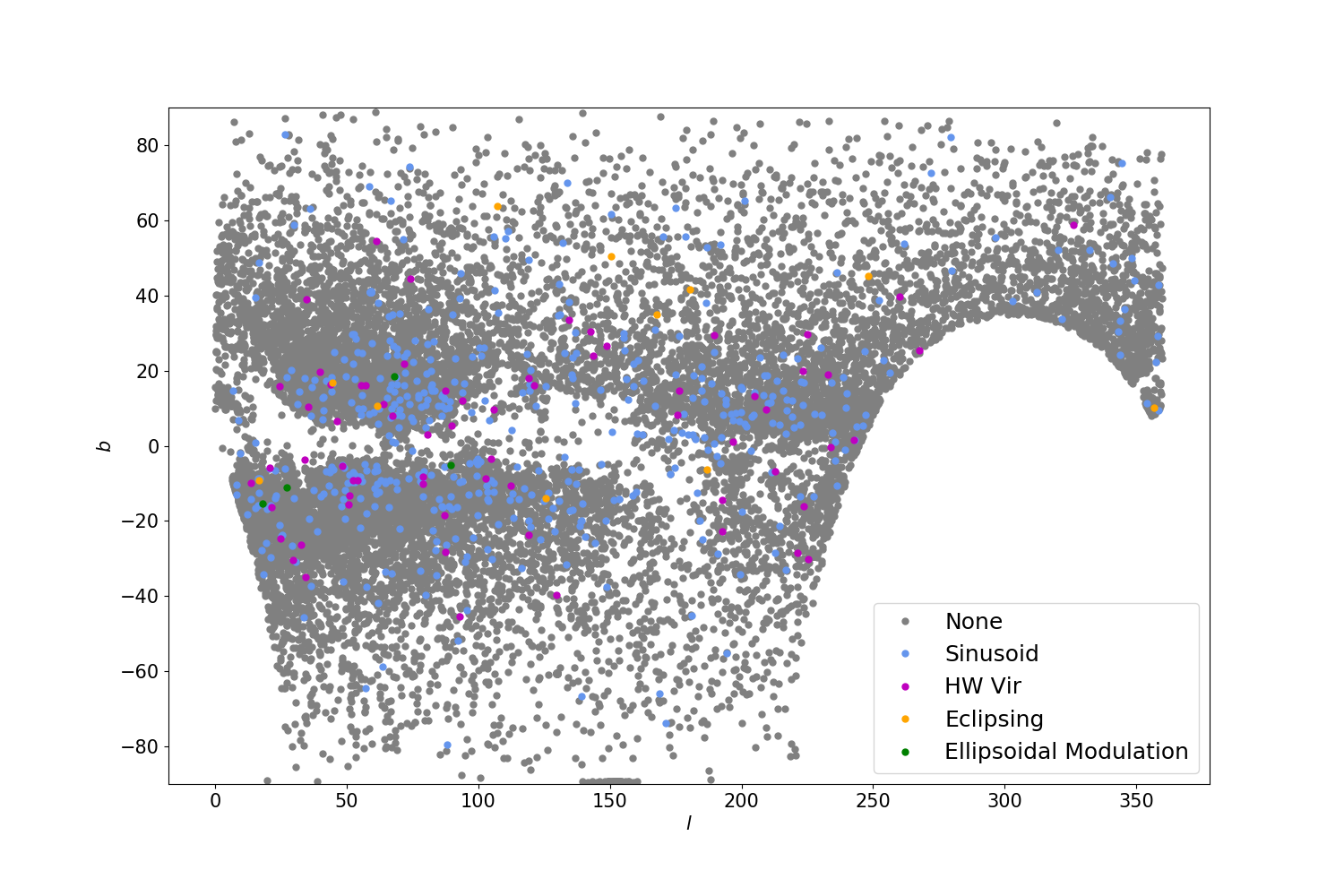}
    \caption{A sky plot of the 18,453 hot subdwarf candidates. The plot shows that the number of variable stars closer to the Galactic Plane is greater compared to regions at higher galactic latitudes.}
    \label{fig:sky}
\end{figure*}

\begin{figure*}
    \centering
    \includegraphics[width=\textwidth]{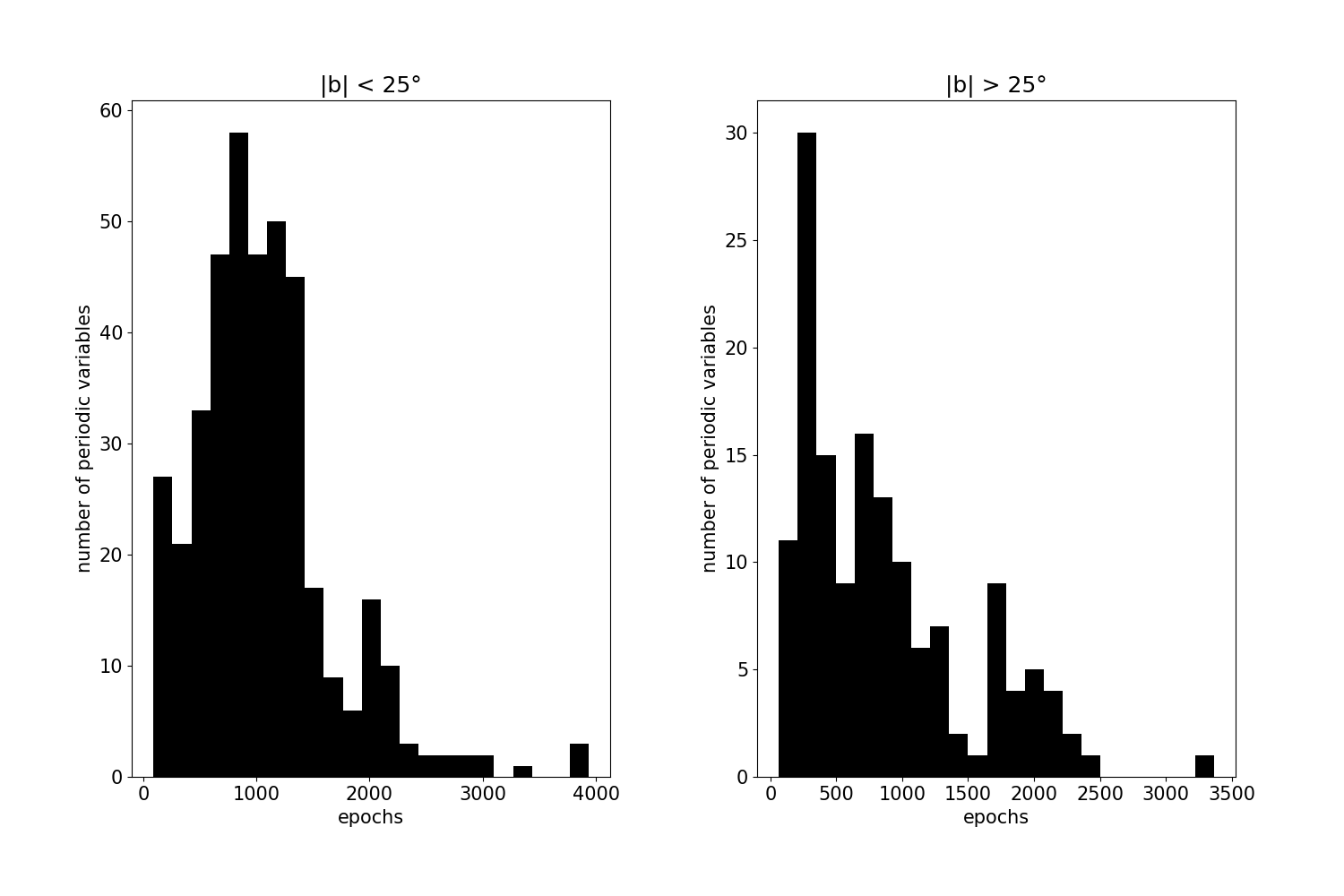}
    \caption{The number of epochs of the ZTF light curves for detected periodic variables near vs. far from the Galactic Plane ($|b|<25^\circ$ vs. $|b|>25^\circ$). The histograms do not show a large difference in epochs, and the substantial number of epochs for both $|b|<25^\circ$ and $|b|>25^\circ$ suggests that the number of epochs is not a signficant explaining factor for the discrepancy of near vs. far from the Galactic Plane illustrated in Fig.\,\ref{fig:sky}.}
    \label{fig:epochs_comp}
\end{figure*}

\subsection{Comparison to the \varindex\, method}

\citet{gui21} developed a method to extract variable stars from a population of sources which uses the empirical Gaia photometric variability metric as a function of magnitude, such as one displayed in Fig.\,\ref{fig:var}. In Gaia photometric uncertainties are empirically determined based on the scatter of individual $G$ magnitude measurements. That means that astrophysical variable stars should have a larger scatter than predicted for a constant star with a comparable magnitude. See \citet{gui21} and \citet{bar22} for more details. Both applied this method to study the variability of white dwarf sources in ZTF \citep{gui21} as well as variable sdB candidates in TESS \citep{bar22}. 

The left panel in Fig.\,\ref{fig:var} displays the empirical Gaia variability for all 18,453 hot subdwarfs along with their classifications to examine the sensitivity of this classification method. Generally, as the magnitude increases, the standard deviation of the flux increases for constant sources, as the measurement of the flux becomes less precise. Accounting for this effect, the main body of constant sources are fitted. We produce a low-degree polynomial fit after $4$x iterative outlier removal, as shown by the bottom black dotted line. The variability index (\varindex) is defined as the residuals for each object after subtracting the best--fitting polynomial fit. Typically sources with \varindex>0.02 are considered variable. The black dashed line in the right panel in Fig.\,\ref{fig:var} represents the variability cutoff for 0.02 above the fit. As shown by the plot, there are a large number of variable sources below the cutoff. Specifically, we calculate that 348 out of 578, or 60.2\%, of all our periodically variable sources would be missed by the 0.02 \varindex\,cut which is dominated by the sinusoidal sources. Only a small fraction from the other three categories will be missed with the 0.02 \varindex\,cut. Categorizing by classification, the following were missed by the cut: 331 out of 496 sinusoids, only 14 out of 67 HW Virs, and only 3 out of 11 eclipsing systems. No ellipsoidal modulations were missed by the 0.02 threshold. These results offers future studies a sense of the fraction of sources missing due to this cut, and suggests that a more careful classification based on Lomb-Scargle and Box Least Squares may be preferable depending on the situation.

\subsection{Hertzsprung-Russell diagram and significance of reddening effect}

We plot all hot subdwarf candidates on the H-R diagram shown in Fig.\,\ref{fig:hr}, only choosing sources with parallax precision better than 10 percent. Additionally, we also show a selection of 1.2 million sources from Gaia eDR3 with parallax precision better than 1 percent as the underlying HR diagram population. These are shown in a grayscale heatmap in log scale.

We noticed that the hot subdwarf candidates we classified as variable are largely concentrated at around 4 to 5 absolute magnitude, which is consistent with previous studies \citep{gei19, gei20}. However, we notice a spread towards redder colors and fainter magnitudes. The spread can be explained with a reddening effect coming from Galactic extinction. Using the Baystar2019 Dustmap \citep{gre19}, we estimated the reddening of hot subdwarf candidates as shown in Fig.\,\ref{fig:reddening-heat}. We find that the objects closer towards the Galactic plane show larger extinction coefficients and therefore suffer from stronger reddening effects. Approximately a fifth of shown variable stars having $G_{BP} - G_{RP} > 0.25$, with a few approaching the outskirts of the main sequence. These results suggest that there could be heavily reddened variable hot subdwarfs hidden within the main sequence stripe and a careful color correction is required to capture heavily reddened hot subdwarfs. All eclipsing sources are found red of the main hot subdwarf clump confirming that they are most likely cataclysmic variable stars. 
%We also find a few sinusoidal variables close to the main sequence populated by B-type stars, indicating that our sample consists of B-

\subsection{Abundance of variable hot subdwarf candidates surrounding the Galactic Plane}

Next, we examine the location of the hot subdwarf candidates in Galactic coordinates, shown in Fig.\,\ref{fig:sky}. Because \citet{gei19} classified hot subdwarfs through conservative quality cuts, most sources in the Galactic Plane were removed because of issues caused by the clustered field and reddening. Thus, the Galactic Plane, represented by $b = 0$ shows a low-density strip. We noticed, however, that most of our periodic variables were clustered around the Galactic Plane---73 percent of our variable sources fall between $-25^\circ < b < 25^\circ$. 

Because there are more overall stars close to the Galactic Plane than the older high-galactic latitude stars farther away, we examined the proportion of variable subdwarfs close and far to the Galactic Plane. We found that 3.6 percent of the hot subdwarf candidates within $25^\circ$ of $b = 0$ are variable, while only 2.0 percent are variable for hot subdwarf candidates at higher galactic latitudes $|b|>25^\circ$. We employed a z-test as shown in equation\,\ref{equ:ztest}  
\begin{equation}\label{equ:ztest}
Z = \frac{p_1-p_2}{\sqrt{p(1-p)(\frac{1}{n_1} + \frac{1}{n_2})}}
\end{equation}

where $n_1$ and $n_2$ are the sample sizes, $p_1$ and $p_2$ are the sample proportions, and $p$ is the total pooled proportion:

\begin{equation}
p = \frac{p_1 n_1 + p_2 n_2}{n_1 + n_2}
\end{equation}

Using the z-test we verify that the difference in density is statistically significant $(p < 0.0001)$. This suggests that more variable hot subdwarf candidates form closer to the Galactic Plane region compared to regions further away from the Galactic Plane. We verify that this effect is not as a result of a discrepancy in the number of epochs near vs. far from the Galactic Plane, which would lead to an observational bias in our ability to detect periodic variables. Thus, we examine the difference of the number of epochs of periodic variables near vs. far from the Galactic Plane in Fig.\,\ref{fig:epochs_comp}, showing no significant variation in the number of epochs. Because the median number of epochs in both cases is large—982 for $|b|<25^\circ$ and 732.5 for $|b|>25^\circ$—we posit that the statistically significant difference in density is not due to the number of epochs. Additionally, the lower Galactic latitude fields are covered by the Galactic Plane survey with a 1-day cadence whereas as most of the higher Galactic latitude sources are covered by the general 3-day cadence observations. \citet{gra19} showed that for different surveys with different cadences the recovery rate flattens out above 200-250 epochs. The median number of epochs for our sources is $>700$ epochs and therefore the slightly different cadence cannot explain the difference in variable sources.

\section{Conclusions and Summary}

We conducted a search for periodic variable hot subdwarf star candidates from the catalogue from \citet{gei19} using the data from ZTF DR6. After converting ZTF-$r$ and ZTF-$g$ magnitudes to normalized flux, we apply Lomb-Scargle and Box Least Squares on all 18,453 hot subdwarf candidates which have a ZTF light curve with at least 30 data points in ZTF-$r$ and ZTF-$g$ combined. We combine manual inspection of the light curves sorted by maximum Lomb-Scargle score with SDE cuts to confidently identify 578 periodic variables. These include 67 HW Vir binaries, 496 reflection effect, pulsator or rotator sinusoidal systems, 11 eclipsing events which are likely cataclysmic variables, and 4 systems which show strong ellipsoidal modulations. Of these, 44 HW Vir binaries, 435 reflection effect, rotation and pulsator sinusoids, 5 eclipsing events, and 2 ellipsoidal modulations have not been previously published. We note that some of the variables are quite bright (up to 13.5\,mag) and well-suited for detailed follow-up investigations, even with smaller-scale facilities. %Only about a dozen HW Virs and a handful of ellipsoidal binaries have been published up to date.

After examining the period distributions of the periodic variables, we applied them to Gaia variability metrics and find that $\approx60$\,\%\, of variable star systems are missed by a 0.02 cut based on the \varindex\,in the Gaia dataset. The majority showing sinusoidal variability. Eclipsing cataclysmic variables, HW Vir binaries and sdBs with strong ellipsoidal modulation are less affected by the 0.02 \varindex\, cut. In the H-R diagram we find that the majority of sources are concentrated around the sdB clump as expected but the data also shows a significant reddening effect, with variable hot subdwarf candidates reaching close to the main sequence stripe in the H-R diagram. Finally, we see almost twice the density of variable stars close to the Galactic Plane than among the high-galactic latitude stars, suggesting that the Galactic Plane is worthwhile to search for new periodically variable subdwarf candidates, provided that light curve pre-processing challenges are addressed.

\section*{Acknowledgements}

We thank the Clark scholar program for their generous support which makes that project possible.

TK and BB acknowledge support from the National Science Foundation through grants AST \#2107982 and AST \#1812874, respectively. TK acknowledges support from NASA through grant 80NSSC22K0338 and from STScI through grant HST-GO-16659.002-A.

Based on observations obtained with the Samuel Oschin Telescope 48-inch at the Palomar Observatory as part of the Zwicky Transient Facility project. ZTF is supported by the National Science Foundation under Grant No. AST-1440341 and a collaboration including Caltech, IPAC, the Weizmann Institute for Science, the Oskar Klein Center at Stockholm University, the University of Maryland, the University of Washington, Deutsches Elektronen-Synchrotron and Humboldt University, Los Alamos National Laboratories, the TANGO Consortium of Taiwan, the University of Wisconsin at Milwaukee, and Lawrence Berkeley National Laboratories. Operations are conducted by COO, IPAC, and UW.

This work has made use of data from the European Space Agency (ESA) mission {\it Gaia} (\url{https://www.cosmos.esa.int/gaia}), processed by the {\it Gaia} Data Processing and Analysis Consortium (DPAC,
\url{https://www.cosmos.esa.int/web/gaia/dpac/consortium}). Funding for the DPAC has been provided by national institutions, in particular the institutions participating in the {\it Gaia} Multilateral Agreement.

The ztfquery code was funded by the European Research Council (ERC) under the European Union's Horizon 2020 research and innovation programme (grant agreement n°759194 - USNAC, PI: Rigault).

\section*{Data Availability}
The ZTF light curves used in this study can be accessed from \url{https://www.ztf.caltech.edu/page/dr6}. A table with all confirmed physically periodic sources will be provided online.
%\end{acknowledgments}

\bibliography{refs}
\bibliographystyle{mnras}

% Don't change these lines
\bsp	% typesetting comment
\label{lastpage}

\end{document}